# Majorana Zero-modes and Topological Phases of Multi-flavored Jackiw-Rebbi model

Shih-Hao Ho[a],[1] Feng-Li Lin[bc],[2,3] and Xiao-Gang Wen[d2]

[1]*Center for Theoretical Physics,*

*Massachusetts Institute of Technology, Cambridge, MA 02139, USA*

[2]*Department of Physics, Massachusetts Institute of Technology,*

*Cambridge, Massachusetts 02139, USA*

[3]*Department of Physics, National Taiwan Normal University, Taipei, 116, Taiwan*

## Abstract

Motivated by the recent Kitaev's K-theory analysis of topological insulators and superconductors, we adopt the same framework to study the topological phase structure of Jackiw-Rebbi model in 3+1 dimensions. According to the K-theory analysis based on the properties of the charge conjugation and time reversal symmetries, we classify the topological phases of the model. In particular, we find that there exist **Z** Majorana zero-modes hosted by the hedgehogs/t'Hooft-Polyakov monopoles, if the model has a $T^2 = 1$ time reversal symmetry. Guided by the K-theory results, we then explicitly show that a single Majorana zero mode solution exists for the SU(2) doublet fermions in some co-dimensional one planes of the mass parameter space. It turns out we can see the existence of none or a single zero mode when the fermion doublet is only two. We then take a step further to consider four-fermion case and find there can be zero, one or two normalizable zero mode in some particular choices of mass matrices. Our results also indicate that a single normalizable Majorana zero mode can be compatible with the cancellation of SU(2) Witten anomaly.

---

[a] shho@mit.edu

[b] linfengli@phy.ntnu.edu.tw

[c] On leave from National Taiwan Normal University.

[d] wen@dao.mit.edu

# CONTENTS

## I. INTRODUCTION

In the recent years, Majorana zero-modes in condensed matter systems have attracted much attention due to its topological and/or symmetry protected degeneracy (similar to the topological ground state degeneracy of topological ordered states). The degeneracy leads to topological and/or symmetry protected non-Abelian geometric phases (up to an overall phase) as we braid the defects that carry Majorana zero-modes. Note that the Hilbert space of these zero-modes are highly entangled [1], implying the underlying topological nature against local perturbations (that may repect certain symmetry in the symmetry protected cases). Thus, these Majorana zero-modes, if exist, can help to realize the robust quantum computation through protected non-Abelian geometric phases. (We would like to point out that those protected non-Abelian geometric phases do not correspond to the standard non-Abelian statistics. This is because the defects that carry Majorana zero-modes have a long range interaction and the overall phase of the non-Abelian geometric phases may depend the exchange paths. So in this paper, we will call such non-Abelian geometric phases, defined up to an overall phase, projective non-Abelian statistics.) .

These fermionic zero-modes appear near the core of some topological defects, which preserve either charge conjugation (particle-hole) symmetry or time reversal symmetry. Especially, such objects in 3+1 dimensions have been proposed by Teo and Kane in [2]. It then opens the possibility to realize the non-trivial projective non-Abelian statistics, the so called projective ribbon statistic in [3] in 3+1 dimensions. Note that, unlike the standard non-Abelian statistics, the non-trivial projective non-Abelian statistics can exist beyond 2+1 dimensions. Moreover, the more general classifications of all possible Majorana zero-modes in the disordered systems of any dimensions has been done in [3–9] based on K-theory analysis in the context of the topological insulators and superconductors [10–15]. The above mentioned 3+1 dimensional Majorana zero modes are one of the examples in this general classification schemes.

On the other hand, in the community of high energy physics the relations between fermionic zero modes and topological defects have been well-studied, and it results in the connection with the index theorem [16]. The pioneer works relevant for our discussions here are the Dirac zero modes around 3+1 dimensional monopole (or hedgehog) by Jackiw and Rebbi [17] and around 2+1 dimensional vortex by Jackiw and Rossi [18]. The zero modes of Majorana fermions, however, are only found in the cases of a kink in one dimension [17] and a vortex background in two dimensions [18]. It is then interesting to see if there exist possible Majorana zero modes in the Jackiw-Rebbi models, which are similar to the ones in Jackiw-Rossi model, as long as the particle-hole symmetry or the time reversal symmetry are preserved by the background topological defects. Once this is possible, it is also interesting to explore the possible connection between the models in high energy physics and the models for topological insulators and superconductors.

A general consideration along the aforementioned direction was first given by Hořava in [19], trying to link the stability of Fermi liquids to the one of D-branes. Some works along this direction in a more specific context have been done in [20–22]. In these works, the relation between Jackiw-Rebbi or Jackiw-Rossi models and the Bogoliubov-de Gennes (BdG) model for topological superconductor is explored and explicitly constructed. However, the role of the SU(2) internal symmetry in the Jackiw-Rebbi model is not fully addressed in these works because the corresponding internal symmetry is not obviously implemented in the BdG model. In this paper, we will further explore this issue by studying the role of SU(2) symmetry in the charge conjugation and time reversal symmetry operations. In this way, we can find out the topological nature of the fermionic zero modes in the framework of K-theory analysis, and then explore the topological phase structures.

In general, the topological defects such as monopoles can be realized and classified by the pattern of spontaneous symmetry breaking. For example, the monopole in the Jackiw-Rebbi model is the 't Hooft-Polyakov one, in which the SU(2) gauge symmetry is broken to U(1) by the Higgs vev of the hedgehog pattern. Thus, the Majorana zero modes, if exist, can be understood as the low energy phenomenon due to the background of topological defects. One would hope that the setting of the theory is also well-defined at high energy regime. This concern has been recently raised by McGreevy and Swingle in [23] for the two-flavored Jackiw-Rebbi model with nonzero Dirac masses. In this case, the theory of odd number of Weyl fermions will be plagued by the Witten anomaly [24]. However, as shown in [23] it is usually difficult to obtain odd number of normalizable Majorana zero modes out of even number Weyl fermions so that the expected projective non-Abelian statistics is in vain. One naive way to bypass this no-go is simply not gauging the internal SU(2) symmetry. Though the presence of the normalizable Majorana zero modes will not be affected by ungauging the internal symmetry, the relative motion of the hedgehog defects on which the Majorana zero modes live will require infinite energy. The reason is that without the cancellation from the gauge fields, the potential between the defects due to the Higgs scalars is confining. If so, these Majorana zero modes will be non-dynamical, not mentioning the nontrivial statistics.

It seems that it is impossible to lift the above no-go result for the 3+1 dimensional Majorana zero modes with nontrivial statistics. However, we will see that in the analysis of [23] the charge conjugation and time reversal symmetries are not imposed. In this paper we will re-examine the issue by requiring the charge conjugation and time reversal symmetries, which then restrict the flavor mass matrices. It turns out that these discrete symmetries are essential to have Majorana zero modes classified by integer **Z** according to the K-theory analysis. Indeed, we find that the no-go theorem is lifted by the existence of $0, 1, 2$ normalizable Majorana zero modes for some ranges of mass parameters in the 4-flavored Jackiw-Rebbi model, which is UV-complete without Witten anomaly.

In general, the symmetry of the point defects determine the possible structures of the zero-modes on the defect. In this paper, we consider the defects that have a time reversal symmetry (generated by $T$) and $Z_2$ symmetry (generated by $C$). The $Z_2$ symmetry can be realized as a charge conjugation symmetry. If $T$ and $C$ satisfy $T^2 = C^2 = 1$ and $TC = CT$, the symmetry group of the defect is given by $G^+_{++}(T,C)$ (using the notation of [7]). From the table VII in [7], we see that the different structures of the zero-modes are classified by **Z**. In this paper, we show that the different structures simply correspond to the number of Majorana zero-modes.

If instead $T$ and $C$ satisfy $T^2 = (-)^{N_F}$, $C^2 = 1$ and $TC = CT$ (where $N_F$ is the total fermion number operator), then the symmetry group of the defect is given by $G^+_{-+}(T,C)$. From the table VII in [7], we find that the different structures of the zero-modes are classified by **Z₂**, corresponding to even or odd numbers of Majorana zero-modes.

Our paper will be organized as follows. In the next section we introduce the Jackiw-Rebbi model with many fermion species/flavors and their charge conjugation (C), time-reversal (T) and parity (P) symmetries. These symmetries also provide a starting point for the review of K-theory analysis on classifications of the topological insulators/superconductors in section III. In section IV we solve the Majorana zero modes for the SU(2) doublet fermions directly from the equations of motion with two and four flavors. We find that there exists a single normalizable Majorana zero mode in both cases. This lifts the no-go result discussed in [23]. We also delineate the phase boundary in the mass parameter space. Finally we conclude with some comments in section V. Technical details about the general choices of the mass matrices's parametrization in four-flavor case will be given in Appendix.

## II. THE JACKIW-REBBI MODEL WITH MULTI-FLAVORED FERMIONS

In this section we will introduce the models for classifying the topologically ordered phases in the context of high energy physics, and also later for finding the Majorana zero modes localized on the point defect. These Majorana zero modes are argued to obey the non-Abelian statistics [3, 23]. The classification of the topologically ordered phases is relevant to finding the fermionic zero modes on the point defects, as suggested by the bulk/edge correspondence of the topological insulators/superconductors [15]. This correspondence relates the pattern of the edge gapless excitations of a topological insulator/superconductor to the one of the bulk gapped excitations, for more conceptual discussion see [25].

## A. The Jackiw-Rebbi model and the discrete symmetries

As shown for examples in [5–8, 25], the classifications of the topological insulators or superconductors are tied to the discrete symmetries of the fermionic systems such as the charge conjugation (C), parity (P) and time reversal (T) symmetry. To further discuss the role of discrete symmetries and be more specific, we first introduce the Jackiw-Rebbi model, which describes Dirac fermions coupled to a gauge field and a scalar field in the SU(2) fundamental/adjoint representation. The Lagrangian density is

$$\mathcal{L} = \bar{\psi}_I i\gamma^\mu D_\mu \psi_I - \mathbf{\Lambda}_{IJ}\bar{\psi}_I \Phi^a T^a \psi_J - \mathbf{m}_{IJ}\bar{\psi}_I \psi_J \;, \tag{2.1}$$

where $D_\mu \equiv \partial_\mu - igA_\mu^a T^a$ is the covariant derivative with $A_\mu^a$ the gauge field, $T^a$'s are the SU(2) generators and $\Phi^a$'s are the scalar fields. The space-time indices are denoted by $\mu, \nu = 0, 1, 2, 3$, the SU(2) indices by $a, b = 1, 2, 3$ and the flavor indices by $I, J = 1, 2, \cdots, N$. The flavor matrices $\mathbf{\Lambda}$ and $\mathbf{m}$ are hermitian. Our convention for gamma matrices is the same as the one in [26], and we will show how to transform this chiral representation to a real representation of gamma matrices later. In this study, we are interested mainly in classifying the bulk topologically ordered phases and then use the results as the guidance in searching for the normalizable fermionic zero modes on the point defect. Since we are only interested in the pattern of these zero modes not their relative dynamics, we will then ignore the gauge fields which play no role for our consideration.

The Hamiltonian derived from the above Lagrangian (without gauge field) is

$$\hat{H} \equiv \psi^\dagger h \psi = \psi_{nI}^\dagger \left(-i\alpha^k \partial_k \otimes \mathbf{1}_{nm} \otimes \mathbf{1}_{IJ} + \beta \otimes \Phi^a T^a_{nm} \otimes \mathbf{\Lambda}_{IJ} + \beta \otimes \mathbf{1}_{nm} \otimes \mathbf{m}_{IJ}\right)\psi_{mJ} \;,$$
$$\equiv \psi^\dagger (-i)\left(\alpha^k \partial_k + \mathbf{M}\right)\psi \;, \tag{2.2}$$

where $\beta \equiv \gamma^0$, $\alpha^k \equiv \gamma^0\gamma^k$ for $k = 1, 2, 3$. This form of the Hamiltonian can be understood as an effective Hamiltonian near the Dirac cone of some electron systems in condensed matters. The constant mass matrix $\mathbf{M}$ is used to gap the system to obtain topological insulators/superconductors provided that there are no zero eigenvalues for $\mathbf{M}$. If the scalar $\Phi^a$ is topologically non-trivial, there may exist some normalizable zero mode solutions which satisfy:

$$h(x)\psi_0(x) = 0. \tag{2.3}$$

Later, we will first classify the these zero modes by K-theory analysis and then use the results as the guidance to solve (2.3) for the zero modes explicitly.

To be specific, in this paper we will consider a particular kind of scalar condensate, the hedgehog configurations as considered in the Jackiw-Rebbi model [17, 23], for which $\Phi^a(r) = \hat{r}^a \phi(r)$ [1]. The core of this point-like defect may host various patterns of normalizable fermionic zero modes depending on what kind of the discrete symmetries or some U(1) symmetries are preserved by the Hamiltonian (2.2).

Three discrete symmetries can be realized in the Jackiw-Rebbi model: parity **P**, charge conjugation **C** and time-reversal symmetry **T** [2]. They are defined by (up to a phase)

$$\hat{P}\hat{\psi}(t,\vec{x})\hat{P}^{-1} = U_p \hat{\psi}(t,-\vec{x}), \tag{2.4}$$

$$\hat{C}\hat{\psi}(t,\vec{x})\hat{C}^{-1} = U_c \hat{\psi}^*(t,\vec{x}), \tag{2.5}$$

$$\hat{T}\hat{\psi}(t,\vec{x})\hat{T}^{-1} = U_t \hat{\psi}(-t,\vec{x}). \tag{2.6}$$

The $U_c$, $U_t$, and $U_p$ are unitary matrices, which can be written in the chiral representation as following:

- for SU(2) doublet, i.e., $T^a = \frac{1}{2}\tau^a$,

$$U_c \equiv (-i\gamma^2 \gamma^5) \otimes (i\tau^2) \otimes \mathbf{1}, \tag{2.7a}$$

$$U_t \equiv (\gamma^1 \gamma^3 \gamma^5) \otimes (i\tau^2) \otimes \mathbf{1}, \tag{2.7b}$$

$$U_p \equiv (\gamma^0) \otimes \mathbf{1} \otimes \mathbf{1}, \tag{2.7c}$$

where $\tau^a$'s are Pauli matrices;

- for SU(2) triplet, i.e., $(T^a)_{nm} = i\epsilon_{nam}$,

$$U_c \equiv (-i\gamma^2) \otimes \mathbf{1} \otimes \mathbf{1}, \tag{2.8a}$$

$$U_t \equiv (\gamma^1 \gamma^3) \otimes \mathbf{1} \otimes \mathbf{1}, \tag{2.8b}$$

$$U_p \equiv (\gamma^0) \otimes \mathbf{1} \otimes \mathbf{1}. \tag{2.8c}$$

---

[1] A typical profile for $\phi(r)$ is $\phi(r) = v \tanh(k_0 r)$ with a constant parameter $k_0$ and the vacuum value of scalar field $v$. We will use this profile to perform the numerical calculations.

[2] Again, we assume these symmetry transformations do not involve the flavor space.

The invariance of the Hamiltonian (2.2) under these transformations give

$$U_c^T h^* = -h U_c^T, \tag{2.9a}$$

$$U_t^\dagger h^* = h U_t^\dagger, \tag{2.9b}$$

$$U_p h = h U_p, \tag{2.9c}$$

respectively. From (2.7), (2.8) and (2.9), the mass matrices $\mathbf{\Lambda}$ and $\mathbf{m}$ have to take the form:

$$\mathbf{\Lambda}^* = \mathbf{\Lambda}, \ \mathbf{m}^* = -\mathbf{m} \tag{2.10a}$$

for SU(2) doublet, and

$$\mathbf{\Lambda}^* = -\mathbf{\Lambda}, \ \mathbf{m}^* = \mathbf{m} \tag{2.10b}$$

for SU(2) triplet. On the other hand, as considering a real (Majorana) fermion, the Dirac spinor is subjected to the reality condition:

$$\hat{\psi}^c(t,\vec{x}) = U_c \hat{\psi}^*(t,\vec{x}) = \hat{\psi}(t,\vec{x}). \tag{2.11}$$

The explicit form of $U_c$ in (2.7a) (also (2.8a)) will be needed later when solving the Majorana zero modes on the point defect, and the parity $s_T \equiv U_t U_t^*$ will be relevant for classifying Majorana zero modes on the point defect based on K-theory analysis.

Some remarks are in order. Firstly, we shall emphasis that the specific form (2.2), rather than the general form commonly considered in [6, 7], is governed by the Lorentz invariance. The special structure of Dirac gamma matrices then assures that parity symmetry is given by (2.7c) (also (2.8c)) since the scalar is treated as a background which does not transform at all under symmetry operations. We also assume CTP theorem so that if $U_c$ is also specified, then $U_t$ will be fixed accordingly.

Secondly, the fermions considered here are fundamental particles, rather than the composite ones with non-trivial constituents or spin-orbital interaction which are commonly realized in condensed matter physics. The lack of nontrivial internal structure for $\psi$ then demands that $U_c U_c^* = +1$ if we would like to obtain a real fermion from a complex one with the supplementary condition (2.11), which indicates the importance of charge conjugation symmetry. This can be easily verified by observing $(\psi^c)^c = U_c^* U_c \psi = \psi$, which implies $U_c U_c^* = +1$. Our choices of $U_c$ in (2.7a) and (2.8a) are conformed to this condition.

Thirdly, suppose we are in a real representation in which the condition (2.11) becomes $\tilde{\psi} = \tilde{\psi}^*$. This implies that the Hamiltonian $\tilde{h}$ is anti-symmetric and purely imaginary by Fermi statistics and hermiticity of $\tilde{h}$. This condition and (2.10) require that

$$\tilde{\gamma}^\mu = \tilde{\gamma}^{\mu*}, \qquad \widetilde{\gamma^0 \tau^a}^* = -\widetilde{\gamma^0 \tau^a} \tag{2.12a}$$

for SU(2) doublet, and

$$\tilde{\gamma}^\mu = -\tilde{\gamma}^{\mu*}, \qquad \widetilde{\gamma^0 \tau^a}^* = \widetilde{\gamma^0 \tau^a} \tag{2.12b}$$

for SU(2) triplet. We will show how to transform the chiral representation (2.7) into this "tilde" representation in the next subsection.

### B. The real representation gamma matrices and real fermions

When discussing symmetry operators such as (2.7) and (2.8), we use the expression of chiral representation of Dirac gamma matrices. We did this in purpose since the chiral representation gives us a great simplification when it comes to finding the zero mode solutions [17, 23]. It is, however, important to have a real representation (in the literal sense) for these symmetries if we adopt the K-theory to classify the bulk topologically ordered phases and the edge zero (gapless) modes for Jackiw-Rebbi model. In this subsection, we show how to transform the symmetry matrices from chiral (complex) representation into a real representation for SU(2) doublet case. This gives an explicit example for the more general discussion given in [25]. For SU(2) triplet fermions, the explicit construction has been given in [20].

For a SU(2) doublet fermion, the scalar field $\Phi = \Phi^a \tau^a / 2$ is in a pseudo-real representation, i.e., $B\Phi^* B^\dagger = -\Phi$ with the unitary matrix $B$ satisfying $BB^* = -1$. The obvious choice is $B = i\tau^2$ which explains the appearance of $i\tau^2$ in (2.7). In order to impose a consistent reality condition on the spinor in (2.11), however, we also have to choose a pseudo-real representation for the Dirac gamma matrices in which there exists an unitary matrix $A$ with $A(\gamma^\mu)^* A^\dagger = \pm \gamma^\mu$ and $AA^* = -1$. We can then always write $A$ as $i\sigma^2 \otimes \mathbf{1}_{2\times 2}$ and combine $A$ and $B$ by a tensor product:

$$U_c' = i\sigma^2 \otimes \mathbf{1}_{2\times 2} \otimes i\tau^2. \tag{2.13}$$

It is obvious that $U'_c U'^*_c = +1$. For such a symmetric matrix $U'_c$, we can always bring $U'_c$ to a new representation $\tilde{U}_c$ by a Takagi transformation in which $\tilde{U}_c = SU'_c S^T = \mathbf{1}_{8\times 8}$ [25]. The only task is to find the unitary matrices $S$.

Let us consider an unitary matrix $S$:

$$S_{8\times 8} = \begin{pmatrix} P_- & -iP_+ \\ iP_+ & -P_- \end{pmatrix} \tag{2.14a}$$

where $P_\pm \equiv \frac{1}{2}(1 \pm \sigma^2 \tau^2)$. Note $P_\pm^2 = P_\pm$ is a projection operator with the properties $P_+ + P_- = 1$, $P_\pm P_\mp = 0$, $P_\pm^T = P_\pm$ and $P_\pm^\dagger = P_\pm$. It is easy to check $S$ is unitary and $\tilde{U}_c = SU'_c S^T$ indeed gives us an identity matrix, i.e.,

$$\begin{aligned}\tilde{U}_c &= SU'_c S^T \\ &= \begin{pmatrix} P_- & -iP_+ \\ iP_+ & -P_- \end{pmatrix} \begin{pmatrix} (i\sigma^2)(i\tau^2) & 0 \\ 0 & (i\sigma^2)(i\tau^2) \end{pmatrix} \begin{pmatrix} P_- & +iP_+ \\ -iP_+ & -P_- \end{pmatrix} \\ &= \begin{pmatrix} (P_+)^2 + (P_-)^2 & 0 \\ 0 & (P_+)^2 + (P_-)^2 \end{pmatrix} = \mathbf{1}_{8\times 8}.\end{aligned} \tag{2.14b}$$

Before we apply (2.14) to transform our chiral representation (2.7a) to the "combined" Majorana representation which mixes up the Lorentz and SU(2) gauge indices, we have to rotate the chiral representation of Dirac gamma matrices into the form (2.13). This can be easily done by

$$U'_c = VU_c V^T, \text{ with an unitary } V = \tfrac{1}{\sqrt{2}} \begin{pmatrix} 1 & 1 \\ -i & i \end{pmatrix}_{4\times 4_L} \otimes \mathbf{1}_{2\times 2}. \tag{2.15}$$

where the subscript $L$ indicates the matrix only act on Lorentz space. Hence we can transform our chiral representation to a Majorana representation by applying $S$ and $V$ sequentially

$$\tilde{U}_c = (SV)U_c(SV)^T, \qquad \tilde{U}_t = (SV)^* U_t (SV)^\dagger, \tag{2.16a}$$

$$\tilde{\psi} = (SV)\psi, \qquad \tilde{\gamma}^\mu = (SV)\gamma^\mu (SV)^\dagger, \quad \mu = 0,1,2,3. \tag{2.16b}$$

Note that $\tilde{\gamma}^\mu$ and $\widetilde{\gamma^0 \tau^a}$, as promised in (2.12), are purely real and imaginary:

$$\tilde{\gamma}^0 = \begin{pmatrix} -\sigma^2 \tau^2 & 0 \\ 0 & \sigma^2 \tau^2 \end{pmatrix}, \quad \tilde{\gamma}^1 = \begin{pmatrix} i\sigma^3 \tau^2 & 0 \\ 0 & i\sigma^3 \tau^2 \end{pmatrix}, \tag{2.17a}$$

$$\tilde{\gamma}^2 = \begin{pmatrix} 0 & i\sigma^2 \\ i\sigma^2 & 0 \end{pmatrix}, \quad \tilde{\gamma}^3 = \begin{pmatrix} -i\sigma^1 \tau^2 & 0 \\ 0 & -i\sigma^1 \tau^2 \end{pmatrix}, \tag{2.17b}$$

$$\widetilde{\gamma^0 \tau^a} = \frac{1}{2} \begin{pmatrix} -\sigma^2(\tau^2 \tau^a - \tau^2(\tau^a)^*) & -i(\tau^a + (\tau^a)^*) \\ i(\tau^a + (\tau^a)^*) & \sigma^2(\tau^2 \tau^a - \tau^2(\tau^a)^*) \end{pmatrix}. \qquad (2.17c)$$

Also, the reality condition (2.11) becomes $\tilde{\psi}^c = \tilde{\psi}^* = \tilde{\psi}$ in this "tilde" representation. This can be verified as follows. We first write the 4-component Dirac spinor as

$$\psi = \begin{pmatrix} \xi \\ \eta \end{pmatrix}. \qquad (2.18a)$$

By (2.11) we arrive

$$\psi^c = (-i\gamma^2 \gamma^5) \otimes (i\tau^2)\psi^* = \psi = \begin{pmatrix} \xi \\ (i\sigma^2)(i\tau^2)\xi^* \end{pmatrix}. \qquad (2.18b)$$

On the other hand, from (2.16) we have

$$\tilde{\psi} = SV \begin{pmatrix} \xi \\ (i\sigma^2)(i\tau^2)\xi^* \end{pmatrix} \qquad (2.18c)$$

$$= -\sqrt{2} \begin{pmatrix} \sigma^2 \tau^2 (Re\ \xi) \\ Im\ \xi \end{pmatrix}$$

where $Re\ \xi$ and $Im\ \xi$ are real and imaginary part of complex spinor $\xi$. From (2.18c), it is clear that $\tilde{\psi}^* = \tilde{\psi}$. This then completes our tasks in this subsection: seeking an unitary matrix $(SV)$ to transform the gamma matrices to the real representation and the reality condition (2.11) becomes $\tilde{\psi}^c = \tilde{\psi}^* = \tilde{\psi}$. However, since the equation of motion is much more simple and easy to deal with in the chiral representation, we will still work in this convention when solving the zero mode functions in Section IV.

### III.   CLASSIFICATION OF TOPOLOGICAL PHASES OF JACKIW-REBBI MODEL

In this section we will classify the topological phases and the edge zero modes of the multi-flavored Jackiw-Rebbi model introduced above. This classification is based on the scheme of K-theory analysis first done by Kitaev in [6]. However, we will mainly follow the conventions given in [7] for our discussions, in which more general results in the context of condensed matters are considered. On the other hand, the more general results in context of high energy physics and AdS/CFT correspondence, see [25].

The K-theory analysis is used to classify the gapped systems, i.e., det$h \neq 0$, and the basic idea is what types of Clifford algebra can be formed for a given Hamiltonian $h$ of massive free fermions in which all structures are fixed except the mass matrix $\mathbf{M}$. The structure of these Clifford algebras yields a configuration space $\mathcal{M}$ for $\mathbf{M}$, whose homotopy groups $\pi_n(\mathcal{M})$'s reflect the topological pattern of the bulk gapped/edge gapless excitations.

To see the structure of the aforementioned structure of Clifford algebra, we first impose the "flat-band" condition $h^2 = \mathbf{1}$ since only the numbers of positive and negative eigenvalues but not their magnitudes are relevant as far as the topological property is concerned. This flat-band condition for Hamiltonian $h$ then is translated to

$$\{\alpha^i \otimes \mathbf{1}, \alpha^j \otimes \mathbf{1}\} = 2\delta^{ij} \otimes \mathbf{1}, \tag{3.1a}$$

$$\{\alpha^i, \mathbf{M}\} = 0, \tag{3.1b}$$

$$\mathbf{M}^2 = -1 \otimes \mathbf{1}. \tag{3.1c}$$

where the $\mathbf{1}$ here is the unit matrix containing gauge and flavor indices. Here we note that for a relativistic theory (3.1a) and (3.1b) are trivially satisfied owing to Lorentz invariance.

Here an important point should be emphasized. When the flat-band condition $h^2 = \mathbf{1}$ is imposed, the space dependence of $\mathbf{M}$ (the hedgehog here) is turned off. In other words, we only focus on the matrix structure of $\mathbf{M}$ when it comes to the Kitaev's K-theory analysis. By doing this, the bulk/edge correspondence is implied manifestly, i.e., we are classifying the same mass matrix structure for bulk (no defect) and edge (defect present) Hamiltonian $h$. As we will see in the next section, the zero mode indeed lives on the defect while different structures of $\mathbf{M}$ gives us different number of zero modes.

We first consider the classification of the topological phases for the complex fermions. For such cases, the flat-band condition (3.1) is all we need to classify the topologically ordered phases. This is because there is no way to form an enlarged Clifford algebra even if there exist some symmetries as we are going to discuss next. From (3.1), the configuration space $\mathcal{M}$ is determined by the complex Clifford algebra $Cl(3, 1)$ with all $\alpha^i$'s fixed, and is denoted by $\mathcal{M} = C_3$ in the convention of [7]. It then gives a trivial topological fermionic phase since $\pi_0(C_3) = 0$ [6, 7, 25].

Now we turn to the case for the real fermions. Naively, if the discrete symmetries are not involved, then the configuration space is $\mathcal{M} = R_3^0$ which again yields only trivial phase as $\pi_0(R_3^0) = 0$. Here the pair $(p, q)$ is referred to the signature of the Clifford algebra $Cl(p, q)$, and for the results of the homotopy groups of $\pi_n(C_p)$ and $\pi_n(R_p^q)$, please check the summarized tables in [6, 7, 25].

However, in this paper the model we consider does respect the charge conjugation and time reversal symmetries. In such a case, we apply these transformations (2.9) to (2.2) then yields the conditions on $\alpha^i$'s and $\mathbf{M}$ as follows:

$$\alpha^i U_p = -U_p \alpha^i, \tag{3.2a}$$

$$\mathbf{M} U_p = U_p \mathbf{M}, \tag{3.2b}$$

$$\alpha^i U_c = U_c (\alpha^i)^*, \tag{3.2c}$$

$$\mathbf{M} U_c = U_c \mathbf{M}^*, \tag{3.2d}$$

$$(\alpha^i)^* U_t = -U_t \alpha^i, \tag{3.2e}$$

$$\mathbf{M}^* U_t = -U_t \mathbf{M}. \tag{3.2f}$$

where $U_p$, $U_c$ and $U_t$ are unitary matrices.

To find the Clifford algebra for K-theory classification, some discussions are in order. Let us observe that it is impossible for $U_p$ to become one of the generators of an enlarged Clifford algebra since it is only anti-commutative with $\alpha^i$ rather than both $\alpha^i$ and $\mathbf{M}$. Therefore, we will ignore parity hereafter.

There are chances for either $U_c$ or $U_t$ to join (3.1) to form a larger Clifford algebra. The only obstacle is the complex conjugate on $\alpha^i$ and $\mathbf{M}$ as shown in (3.2). However, if there exists real representation for all the generators appear in (3.2), then the resultant Clifford algebra can be enlarged than the one without these symmetries. As shown in II B we indeed can find such a representation for Jackiw-Rebbi model. Moreover, in such representation $U_c = 1$ so that it becomes trivial and plays no role in forming the enlarged Clifford algebra. However, we should caution the readers that there are more general consideration for which $U_c$ cannot be transformed into unity in the context of condensed matter physics.

Moreover, in such a Majorana representation, all the gamma matrices are all either real or pure imaginary so that $\alpha^i$'s, $\mathbf{M}$ and $U_t$ constructed from them will also be all either real. In such a case, $U_t$ then anti-commutes with $\alpha^i$'s and $\mathbf{M}$ as seen from (3.2e) and (3.2f), it then yields an enlarged Clifford algebra. The structure of the resultant Clifford algebra will then only depend on the parity parameter

$$s_T \equiv U_t U_t^* = U_t^2. \tag{3.3}$$

| Fermions | m | Λ | $U_c$ | $U_t$ | $s_T$ | $\mathcal{M}$ | bulk phases | point defect |
|---|---|---|---|---|---|---|---|---|
| Doublet | Im | Re | $(i\gamma^2\gamma^5) \otimes (i\tau^2)$ | $(\gamma^1\gamma^3\gamma^5) \otimes (i\tau^2)$ | +1 | $R_4^0$ | 0 | $\mathbf{Z}$ |
| Triplet | Re | Im | $(-i\gamma^2)$ | $\gamma^1\gamma^3$ | −1 | $R_3^1$ | $\mathbf{Z}$ | $\mathbf{Z}_2$ |

TABLE I. The summary of symmetry operators and topological class for real fermions in Jackiw-Rebbi model.

In the second equality we have assumed $U_t$ to be real, so are $\alpha^i$'s and $\mathbf{M}$ [3].

It is straightforward to show that $s_T$ is invariant under Takagi transformation. Thus it is easy to check that $s_T = 1$ for SU(2) doublet, and $s_T = -1$ for SU(2) triplet. For $s_T = 1$, in the convention of [7] the resultant Clifford algebra is $Cl(4,1)$ so that $\mathcal{M} = R_4^0$ and $\pi_0(R_4^0) = 0$ yields only trivial bulk phase. On the other hand, for $s_T = -1$ the resultant Clifford algebra is $Cl(3,2)$ so that $\mathcal{M} = R_3^1$. Using the result of Bott's periodicity theorem for K-theory, i.e.,

$$R_p^q = R_{p-q \bmod 8}^0, \tag{3.4}$$

$$\pi_n(R_p^0) = \pi_0(R_{p-n}^0)., \tag{3.5}$$

the bulk topological phases for real SU(2) triplet fermions are classified by $\pi_0(R_3^1) = \pi_0(R_2^0) = \mathbf{Z}$.

Moreover, we can also classify the edge Majorana zero modes with the help of (3.5). This is assured by the bulk/edge correspondence for the topological insulators/superconductors, which here is manifested as

$$\pi_{d-d_b-1}(\mathcal{M}) \tag{3.6}$$

for the edge modes living on $d_b$-dimensional defect. We are interested in the Majorana zero modes living on the point defect in this paper, $d_b = 0$ and then the classification is given by $\pi_2(R_4^0) = \mathbf{Z}$ for $s_T = 1$ and $\pi_2(R_3^1) = \mathbf{Z}_2$ for $s_T = -1$. These results are summarized in Table I. Note that if there is no symmetry, the Majorana edge modes on the point defect is classified by $\pi_2(R_3^0) = \mathbf{Z}_2$, however, in this paper we will not solve the zero modes for such a case.

---

[3] Due to the hermiticity of $\alpha^i$ and anti-hermiticity of $\mathbf{M}$, $\alpha^i$'s are symmetric and $\mathbf{M}$ is anti-symmetric if they are all real.

Before we close this section, we should note that the K-theory analysis discussed here can only guide us to find the number of zero modes hosted on the topological defect. However, it cannot tell us the detailed state space structure of these zero modes. This issue has been addressed in 2+1 Jackiw-Rossi model with the background consisting of integer number of vortices in [27]. A general formula for the dimensionality of state space is given explicitly and the mode operators with fermion number conservation are realized by a restricted Clifford algebra which shares some similarities with K-thoery approach.

## IV. MAJORANA ZERO MODES IN TWO- AND FOUR-FLAVORED JACKIW-REBBI MODEL

In this section, we discuss the structure of mass matrices of Majorana fermions in the SU(2) doublet representation in Jackiw-Rebbi model and solve the coupled differential equation numerically. We find that there exists either no normalizable zero mode or a single Majorana zero mode in two-flavor case, which depends on the Dirac mass parameters. When the number of fermion species becomes four, we can find that the number of normalizable zero modes may be zero, one or two which also depends on the Dirac mass parameters.

As we discussed in Section III, the number of normalizable zero modes hosted on the point defect is expected to be the same as the one indicated by K-theory analysis. Hence, the criteria to write down the appropriate mass matrices $\mathbf{m}$ and $\mathbf{\Lambda}$ in question are flat-band condition (3.1c) and charge conjugation symmetry (2.9a), which gives

$$\{\mathbf{\Lambda}, \mathbf{m}\} = 0, \tag{4.1a}$$

$$\mathbf{\Lambda}^2 + \mathbf{m}^2 = \mathbf{1}, \tag{4.1b}$$

$$\mathbf{\Lambda}^* = \mathbf{\Lambda}, \ \mathbf{m}^* = -\mathbf{m}. \tag{4.1c}$$

For the given scalar condensate $\Phi^a(r)$ (a hedgehog here) and proper boundary conditions for the mode function, (4.1) are all we need to find normalizable zero modes indicated by K-theory analysis in Section III.

## A.  Equations of motion for the zero mode

The equation of motion for the Majorana zero modes governed by the Hamiltonian (2.2) is

$$\left(\alpha^k p_k \mathbf{1}_{nm}\mathbf{1}_{IJ} + \beta \mathbf{M}_{nmIJ}\right)\psi_{nJ} = 0, \tag{4.2a}$$

with the reality condition on the Dirac field

$$\psi = \psi^c = U_c\psi^* = \begin{pmatrix} \xi \\ i\sigma^2 i\tau^2 \xi^* \end{pmatrix}. \tag{4.2b}$$

as shown in (2.18b).

To proceed, we notice that the $SU(2)_{Lorentz} \times SU(2)_{internal}$ symmetry has been broken explicitly by the background scalar field to the diagonal $SU(2)$, the subgroup which mixes spatial and internal symmetries [17]. So we decompose the two component spinor $\xi$ as

$$(\xi_{ia})_J = (g_J \delta_{ib} + \vec{g}_J \vec{\sigma}_{ib})(i\tau^2)_{ba} \tag{4.3}$$

where $i, j = 1, 2$ and $a, b = 1, 2$ are SU(2) and spin indices respectively. With this decomposition (4.3), (4.2) becomes a pair coupled differential equations:

$$\partial_r h_I + \mathbf{\Lambda}_{IJ}\frac{\phi(r)}{2}h_J + \mathbf{\Omega}_{IJ}h_{rJ} = 0 \tag{4.4a}$$

$$\partial_r (r^2 h_{rI}) + \mathbf{\Lambda}_{IJ}\frac{\phi(r)}{2}(r^2 h_{rJ}) - \mathbf{\Omega}_{IJ}(r^2 h_J) = 0 \tag{4.4b}$$

where

$$g(r) = e^{+i\pi/4}h(r), \quad \vec{g} = \hat{r}g_r(r) = e^{-i\pi/4}h_r(r)\hat{r} \tag{4.5}$$

and define

$$\mathbf{\Omega} \equiv i\mathbf{m} \tag{4.6}$$

so that $\Omega$ is real for pure imaginary $\mathbf{m}$. We made an ansatz $\vec{g} = \hat{r}g_r(r)$ because of the rotation symmetry of the background field. Note that (4.4) are the differential equations with real coefficients so that we can solve for the "Majorana" zero modes.

In order to find a normal mode, we have to transform the coupled first order differential equations (4.4) into a set of decoupled second order differential equations [4]. First, we define:

---

[4] The derivation given here is closely related to [23] except there it imposes different condition on $\mathbf{m}$ and thus yields different pattern from ours for zero modes of two-flavor case.

$$W_I \equiv (e^{\int^r ds \mathbf{\Lambda}\phi(s)/2})_{IJ}(h_J(r)), \tag{4.7a}$$

$$W_{rI} \equiv (e^{\int^r ds \mathbf{\Lambda}\phi(s)/2})_{IJ}(r^2 h_{rJ}(r)). \tag{4.7b}$$

Hereafter we ignore the flavor indices. Then we can rewrite (4.4b) as

$$\partial_r W_r(r) = e^{\int^r ds \mathbf{\Lambda}\phi(s)/2} \mathbf{\Omega}(r^2 h). \tag{4.8a}$$

Integrating (4.8a) gives

$$r^2 h_r(r) = e^{-\int^r ds \mathbf{\Lambda}\phi(s)/2} \int^r d\tilde{r} e^{\int^{\tilde{r}} ds \mathbf{\Lambda}\phi(s)/2} \mathbf{\Omega}(\tilde{r}^2 h(\tilde{r})). \tag{4.8b}$$

In use of (4.7a) and (4.8b), we can rewrite (4.4a):

$$r^2 e^{-\int^r ds \mathbf{\Lambda}\phi(s)/2} \partial_r W(r) = -\mathbf{\Omega}\ (r^2 h_r)$$

$$= -\mathbf{\Omega}\ e^{-\int^r ds \mathbf{\Lambda}\phi(s)/2} \int^r d\tilde{r} e^{\int^{\tilde{r}} ds \mathbf{\Lambda}\phi(s)/2} \mathbf{\Omega}(\tilde{r}^2 h(\tilde{r}))$$

$$= -e^{+\int^r ds \mathbf{\Lambda}\phi(s)/2} \mathbf{\Omega}^2 \int^r d\tilde{r} e^{-\int^{\tilde{r}} ds \mathbf{\Lambda}\phi(s)/2} (\tilde{r}^2 h(\tilde{r})) \tag{4.9a}$$

where in the third equality we have used the fact $\mathbf{\Omega}\mathbf{\Lambda} + \mathbf{\Lambda}\mathbf{\Omega} = 0$ twice. Since we choose to work in the basis where both $\mathbf{\Lambda}$ and $\mathbf{\Omega}^2$ are diagonal, differentiating (4.9a) arrives a set of decoupled second-order differential equations:

$$\partial_r \left( r^2 e^{-\int^r ds \Lambda_{ii}\phi(s)} \partial_r W(r) \right) = -\Omega_{ii}^2 r^2 e^{-\int^r ds \Lambda_{ii}\phi(s)} W(r) \tag{4.9b}$$

where the $\Lambda_{ii}$ and $\Omega_{ii}$ are diagonal elements of $\mathbf{\Lambda}$ and $\mathbf{\Omega}$. In the following discussion, we will ignore the subscript $i$ and use the unbold letters $\Lambda$ and $m$ to denote one of the diagonal elements for simplicity.

The first thing we notice is that the solution of (4.9b) when the diagonal matrix element $\Omega^2 = 0$ is

$$h(r) = e^{-\int^r ds \Lambda\phi(s)/2} \left( c_2 + \int^r d\tilde{r} e^{+\int^{\tilde{r}} ds \Lambda\phi(s)} \frac{c_1}{\tilde{r}^2} \right). \tag{4.10}$$

Since we require the mode function is regular at the origin and the profile of $\phi(r)$ is $\phi(r = 0) = 0$ and $\phi(r \to \infty) = v$ (we set $v = 1$ throughout the paper for simplicity) [5], the regularity near the origin implies $c_1 = 0$. Therefore, we can see that the solution $h(r)$ is normalizable for $\Lambda > 0$ as $\Omega^2 = 0$.

---

[5] A typical choice of $\phi(r)$ is $\phi(r) = v\ tanh(k_0 r)$ where $k_0$ is a constant.

For diagonal elements $\Omega^2 \neq 0$ ($\Omega_{ii}^2 \equiv -\mathbf{m}_{ii}^2 < 0$ since $\mathbf{\Omega}$ is a real anti-symmetric matrix), the solution for (4.9b) is

$$W(r \to 0) \sim \frac{c}{r}\left(e^{-mr} - e^{+mr}\right) \tag{4.11}$$

because we require $W(r=0)$ is regular and $\phi(r=0) = 0$. For general $r$, the $W(r)$ satisfies the differential equation

$$r^2 W'' + (2r - r^2 \Lambda \phi(r))W' - m^2 r^2 W = 0. \tag{4.12}$$

To derive the proper boundary condition for numerical calculation near $r = 0$, we first expand (4.11) near $r \sim 0$, i.e.,

$$W(r \ll 1) \sim -2mc\left(1 + \frac{m^2 r^2}{3!} + \frac{m^4 r^4}{5!}\right), \tag{4.13a}$$

$$W'(r \ll 1) \sim -\frac{2}{3}m^3 cr\left(1 + \frac{m^2 r^2}{10} + \frac{m^4 r^4}{280}\right). \tag{4.13b}$$

Then we use the polynomial expansion to solve (4.12) near $r = 0$ and the solution should reproduce the result (4.13) when we turn off $\phi(r)$. Here we use the profile of $\phi(r) = tanh(k_0 r)$ where $k_0$ is a constant, and the results are

$$W(r) \sim a_0 + a_1 r + a_2 r^2 + a_3 r^3 + a_4 r^4 + a_5 r^5 + \cdots, \tag{4.14a}$$

$$W'(r) \sim a_1 + 2a_2 r + 3a_3 r^2 + 4a_4 r^3 + 5a_5 r^4 + \cdots, \tag{4.14b}$$

$$\phi(r) \sim k_0 r - \frac{(k_0 r)^3}{3} + \cdots. \tag{4.14c}$$

Substitute (4.14) into (4.12) and compare the coefficients, we arrive

$$r^{-1} \text{ term: } a_1 = 0, \tag{4.15a}$$

$$\text{Constant term: } a_2 = \frac{a_0 m^2}{6}, \tag{4.15b}$$

$$r \text{ term: } a_3 = \frac{\Lambda k_0 + m^2}{12} a_1 = 0, \tag{4.15c}$$

$$r^2 \text{ term: } a_4 = \frac{2\Lambda k_0 + m^2}{20} a_2 = \frac{m^2(2\Lambda k_0 + m^2)}{120} a_0, \tag{4.15d}$$

$$r^3 \text{ term: } a_5 = \frac{3\Lambda k_0 + m^2}{26} a_3 = 0. \tag{4.15e}$$

So we take $a_0 = 1$ and set $c = \frac{-1}{2m}$ to match (4.13). Then the boundary conditions are

$$W(r \to 0) \sim \left(1 + \frac{m^2 r^2}{6} + \frac{m^2(2\Lambda k_0 + m^2)}{120} r^4\right), \tag{4.16a}$$

$$W'(r \to 0) \sim \frac{m^2 r}{3}\left(1 + \frac{2\Lambda k_0 + m^2}{10} r^2\right). \tag{4.16b}$$

Using the boundary condition (4.16), we can then adopt the numerical shooting method to find the normalizable zero mode solutions of the second-order differential equation (4.12).

### B. Two-flavored case

In two-flavor case, we can now parametrize our mass matrices to satisfy (4.1) as

$$\mathbf{\Lambda} = \lambda_1 \sigma^1 + \lambda_2 \sigma^3, \tag{4.17a}$$

$$\mathbf{m} = \mu \sigma^2 \tag{4.17b}$$

where $\lambda_1$, $\lambda_2$ and $\mu$ are real numbers with the supplement condition $\lambda_1^2 + \lambda_2^2 + \mu^2 = 1$. We make a rotation to the basis where $\mathbf{\Lambda}$ is diagonal, i.e.,

$$O = \begin{pmatrix} \frac{\lambda_1}{\sqrt{\lambda_1^2 + (\tilde{\lambda} - \lambda_2)^2}} & \frac{\tilde{\lambda} - \lambda_2}{\sqrt{\lambda_1^2 + (\tilde{\lambda} - \lambda_2)^2}} \\ \frac{\lambda_1}{\sqrt{\lambda_1^2 + (\tilde{\lambda} - \lambda_2)^2}} & \frac{-\lambda_2 - \tilde{\lambda}}{\sqrt{\lambda_1^2 + (\tilde{\lambda} - \lambda_2)^2}} \end{pmatrix}, \tag{4.18a}$$

$$\mathbf{\Lambda}' = O \mathbf{\Lambda} O^T = \begin{pmatrix} \tilde{\lambda} & 0 \\ 0 & -\tilde{\lambda} \end{pmatrix}, \tag{4.18b}$$

$$\mathbf{m}' = O \mathbf{m} O^T = \mathbf{m}\, det(O) = \frac{-\lambda_1}{|\lambda_1|} \mathbf{m} \tag{4.18c}$$

where $\tilde{\lambda} \equiv \sqrt{\lambda_1^2 + \lambda_2^2} > 0$. In Fig. 1 we show the numerical result for the case $\mu \neq 0$, which gives us no normalizable zero energy solution. If $\mu$ is tuned to be zero while keeping $\mathbf{\Lambda}$ the same, a single zero mode solution is popped out because of (4.10). The phase diagram is shown in Fig. 2.

Here we should note a difference between the consideration in [23] and ours. We start with the hermitian mass matrices which are implied by the hermiticity of the Hamiltonian. On the other hand, in [23] they start from a generic two-Weyl-doublet model in which the hermiticity of their Hamiltonian is guaranteed by adding the hermitian conjugate rather than requiring the mass matrices to be hermitian. One may expect their mass matrices are more general than the ones considered here. It is, however, not true because they also require $\mathbf{m}$ to be hermitian after performing a Takagi transformation which makes $\mathbf{\Lambda}$ contain only real and positive diagonal entries. This actually imposes a constraint on the mass matrices and makes their model fall into different class from ours, so that there is no single normalizable zero mode found in their model.

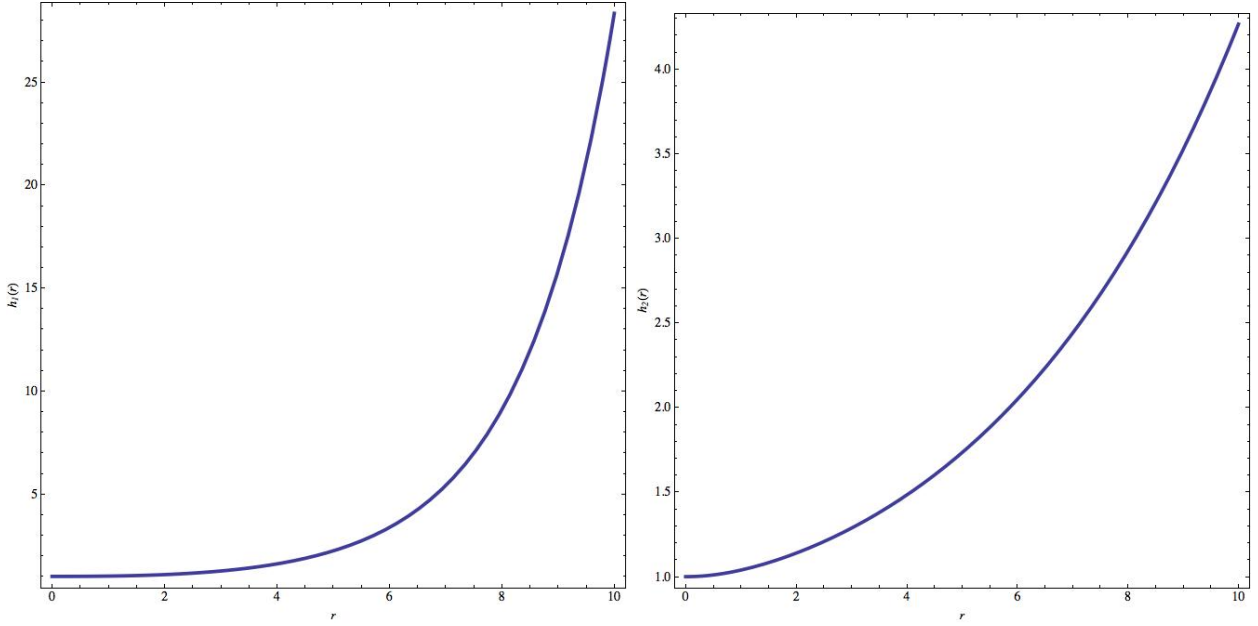

FIG. 1. The profile of zero mode solutions of (4.4) for the mass matrices (4.18). The parameter choices are $\tilde{\lambda} = 0.6$, $\mu = 0.4$ and $k_0 = 1$. Left: $h_1(r)$. Right: $h_2(r)$

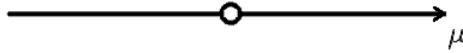

FIG. 2. Phase diagram for (4.4) with mass matrix (4.18) of 2-flavored Majorana fermions. We assume $\tilde{\lambda} > 0$. The normalizability of the zero modes only depends on the Dirac mass $\mu$. The single normalizable zero mode appears only at the origin of this one dimensional phase space.

## C. Four-flavored case

Now we would like to consider the case with four flavors of fermions, and we find that there exists a single normalizable zero mode in this case even for a Dirac mass matrix with some nonzero elements. We parametrize our mass matrices, which satisfy the requirement

(4.1c), by an outer product of two Pauli matrices:

$$\begin{aligned}\boldsymbol{\Lambda} =& \lambda_1(\sigma^0 \otimes \sigma^1) + \lambda_2(\sigma^0 \otimes \sigma^3) + \lambda_3(\sigma^0 \otimes \sigma^0) + \lambda_4(\sigma^2 \otimes \sigma^2) \\ &+ \lambda_5(\sigma^1 \otimes \sigma^0) + \lambda_6(\sigma^1 \otimes \sigma^3) + \lambda_7(\sigma^1 \otimes \sigma^1) + \lambda_8(\sigma^3 \otimes \sigma^0) \\ &+ \lambda_9(\sigma^3 \otimes \sigma^1) + \lambda_{10}(\sigma^3 \otimes \sigma^3),\end{aligned} \quad (4.19\text{a})$$

$$\begin{aligned}\mathbf{m} =& m_1(\sigma^0 \otimes \sigma^2) + m_2(\sigma^1 \otimes \sigma^2) + m_3(\sigma^3 \otimes \sigma^2) \\ &+ m_4(\sigma^2 \otimes \sigma^0) + m_5(\sigma^2 \otimes \sigma^1) + m_6(\sigma^2 \otimes \sigma^3).\end{aligned} \quad (4.19\text{b})$$

In addition, to meet the condition (4.1a) and choose a basis in which $\boldsymbol{\Lambda}$ is diagonal, (4.19) can then be further reduced to

$$\boldsymbol{\Lambda} = \lambda_2(\sigma^0 \otimes \sigma^3) + \lambda_3(\sigma^0 \otimes \sigma^0) + \lambda_8(\sigma^3 \otimes \sigma^0) + \lambda_{10}(\sigma^3 \otimes \sigma^3), \quad (4.20\text{a})$$

$$\begin{aligned}\mathbf{m} =& m_1(\sigma^0 \otimes \sigma^2) + m_2(\sigma^1 \otimes \sigma^2) + m_3(\sigma^3 \otimes \sigma^2) \\ &+ m_4(\sigma^2 \otimes \sigma^0) + m_5(\sigma^2 \otimes \sigma^1) + m_6(\sigma^2 \otimes \sigma^3).\end{aligned} \quad (4.20\text{b})$$

Note that the mass parameter space has been trimmed down largely.

Next we consider one special case of mass matrices (A11b) which is one among three cases in (A11), and leave the detailed discussions about how to derive these proper mass matrices in Appendix A. The reason to choose this particular case for illustration is that the normalizability of the zero modes can be seen from a set of decoupled first-order differential equations rather than a set of second-order ones. For the other two cases in (A11), we have to solve the second order differential equations for the zero modes numerically. The special choices of mass matrices are

$$\boldsymbol{\Lambda} = \lambda_2(\sigma^0 \otimes \sigma^3) + \lambda_8(\sigma^3 \otimes \sigma^0), \quad (4.21\text{a})$$

$$\mathbf{m} = m_2(\sigma^1 \otimes \sigma^2) + m_5(\sigma^2 \otimes \sigma^1). \quad (4.21\text{b})$$

In this basis, $\boldsymbol{\Lambda}$ and $\mathbf{m}$ aren diagonal and totally off-diagonal, respectively. After tedious manipulations, we are able to write down (4.4a) explicitly, i.e.,

$$\partial_r h_1 + (\lambda_2 + \lambda_8)\frac{\phi(r)}{2}h_1 + (m_2 + m_5)h_{r4} = 0, \quad (4.22\text{a})$$

$$\partial_r h_4 - (\lambda_2 + \lambda_8)\frac{\phi(r)}{2}h_4 - (m_2 + m_5)h_{r1} = 0, \quad (4.22\text{b})$$

$$\partial_r h_2 - (\lambda_2 - \lambda_8)\frac{\phi(r)}{2}h_2 - (m_2 - m_5)h_{r3} = 0, \quad (4.22\text{c})$$

$$\partial_r h_3 + (\lambda_2 - \lambda_8)\frac{\phi(r)}{2}h_3 + (m_2 - m_5)h_{r2} = 0. \quad (4.22\text{d})$$

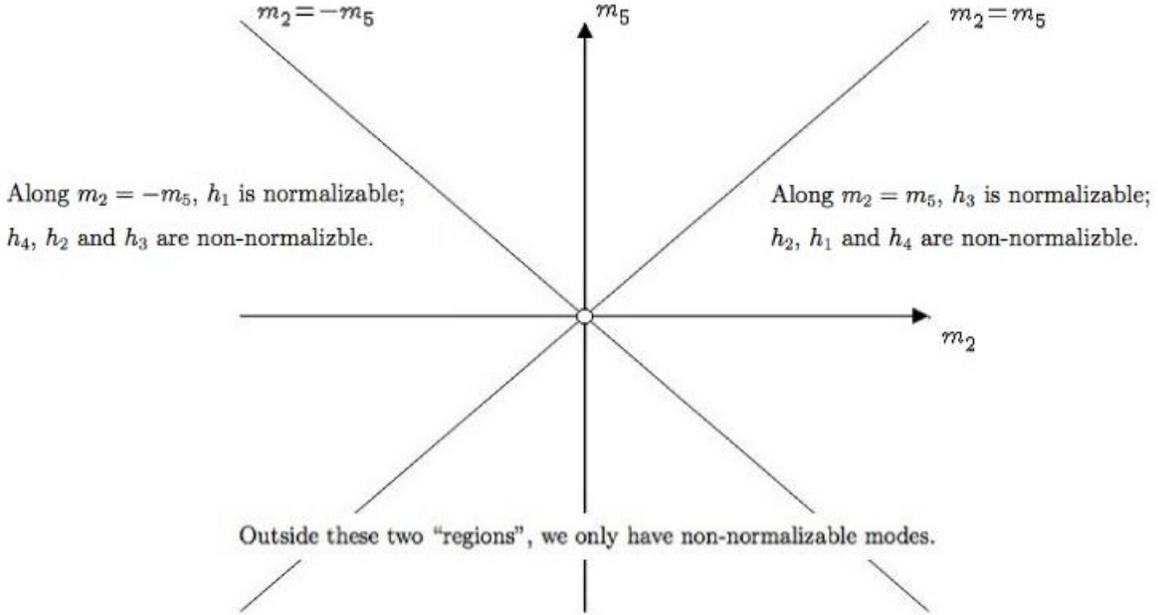

FIG. 3. Phase diagram for (4.22) of 4-flavored Majorana fermions. We assume $\lambda_2 > \lambda_8 > 0$. The two normalizable zero mode solutions appear at the origin. Along two straight lines $m_2 = \pm m_5$, there exists only single normalizable solution. Other than these regions, there exists no normalizable solution.

We can see that (4.22a), (4.22b) and (4.22c), (4.22d) are paired to form two copies of the two-flavor equations. For arbitrary nonzero values of $|\lambda_2| \neq |\lambda_8|$ and $|m_2| \neq |m_8|$, there are no normalizable zero mode solutions in (4.22). However, there is a single zero mode solution when $|m_2| = |m_5|$ ($|\lambda_2| \neq |\lambda_8|$). For $m_2 = m_5$, the pair $h_1$ and $h_4$ do not have normalizable solutions while $h_3$ is normalizable and $h_2$ is non-normalizable if $\lambda_2 - \lambda_8 > 0$ and vice versa. In the case of $m_2 = -m_5$, the pair $h_2$ and $h_3$ do not have normalizable solutions while $h_1$ is normalizable and $h_4$ is non-normalizable if $\lambda_2 + \lambda_8 > 0$ and vice versa. Moreover, two normalizable solutions are present when $m_2 = m_5 = 0$. That is, we have found $0, 1, 2$ normalizable zero modes depending the mass parameters. We show the phase diagram in Fig. 3.

The phase diagram shows that this single Majorana zero mode exists only in a small range of mass parameter space - a co-dimensional one plane; in other words, a single Majorana zero-mode emerges only when some diagonal elements (but not all) of the square of Dirac mass $\mathbf{m}^2$ equal to zero. The two normalizable solutions only appears at the origin of two-dimensional $(m_2, m_5)$ plane, which is similar to two-flavor case: a single zero mode solution appears only at the origin of one-dimensional space $\mu$. This phase diagram can be viewed as a stack of two one-dimensional phase diagrams, each of which corresponds to the straight line $m_2 = \pm m_5$ respectively. This simply reflects the feature of (4.22), which is seen to be two copies of two-flavor equations.

Moreover, the pattern of normalizable zero modes found above is in accordance with the result of K-theory analysis, namely, classified by $\pi_2(R_4^0) = \mathbf{Z}$. In contrast, in the 2-flavored case it is hard to tell if the pattern of the normalizable zero modes is classified by $\mathbf{Z}$ or $\mathbf{Z}_2$. This just reflects the fact that the K-theory classification needs large number of spectators to ensure stable results.

## V. CONCLUSIONS

Conventionally Majorana zero-modes are realized by splitting a complex Dirac fermion into two real degrees of freedom, i.e., the Majorana zero modes are not fundamental and always come in pairs. An attempt to find a single real mode was made by McGreevy and Swingle in [23], in which they try to realize a single normalizable zero mode emerging from two SU(2) doublet Weyl fermions to avoid Witten anomaly. They found that it is impossible to realize a single Majorana zero mode in an Witten anomaly-free theory.

Despite their promising attempt, we turn to the K-theory analysis for help. By adopting flat-band condition and accommodating the discrete symmetries, which are essential elements in K-thoery analysis, we constraint the mass matrix structure in (2.2). Furthermore, we introduce the Majorana fermions by imposing the reality condition on complex Dirac fields, which is commonly used in high energy physics. It turns out that the K-theory analysis tells us that there could be any integer number of normalizable zero modes hosted by the hedgehog (Table I) and indeed we can see the existence of a single zero mode when the fermion doublet is only two. We then take a step further to consider four-fermion case and find there can be zero, one or two normalizable zero mode in some particular choices of mass matrices.

In summary, a single Majorana zero mode can be realized in Jackiw-Rebbi model with some constraints given by the discrete symmetries. We believe there exists an index theorem to support this result since it is a result guided by K-theory. It will be an interesting future project to find the index theorem behind.

**ACKNOWLEDGMENTS**

SHH would like to thank Professor Roman Jackiw for stimulating discussions. FLL is supported by Taiwan's NSC grants (grant NO. 100-2811-M-003-011 and 100-2918-I-003-008). SHH and FLL thank the support of NCTS.

**Appendix A: Four-Flavored Mass Matrices**

With four flavors, we only need to extend our mass matrix to 4x4 matrices, we can also parametrize them by an outter product of two Pauli matrices:

$$\begin{aligned}
\mathbf{\Lambda} =& \lambda_1(\sigma^0 \otimes \sigma^1) + \lambda_2(\sigma^0 \otimes \sigma^3) + \lambda_3(\sigma^0 \otimes \sigma^0) + \lambda_4(\sigma^2 \otimes \sigma^2) \\
& + \lambda_5(\sigma^1 \otimes \sigma^0) + \lambda_6(\sigma^1 \otimes \sigma^3) + \lambda_7(\sigma^1 \otimes \sigma^1) + \lambda_8(\sigma^3 \otimes \sigma^0) \\
& + \lambda_9(\sigma^3 \otimes \sigma^1) + \lambda_{10}(\sigma^3 \otimes \sigma^3), & \text{(A1a)} \\
\mathbf{m} =& m_1(\sigma^0 \otimes \sigma^2) + m_2(\sigma^1 \otimes \sigma^2) + m_3(\sigma^3 \otimes \sigma^2) \\
& + m_4(\sigma^2 \otimes \sigma^0) + m_5(\sigma^2 \otimes \sigma^1) + m_6(\sigma^2 \otimes \sigma^3). & \text{(A1b)}
\end{aligned}$$

Applying constraint (4.1a) leads to six equations:

$$\begin{aligned}
(\sigma^2 \otimes \sigma^1): & \quad \lambda_1 m_4 + \lambda_3 m_5 - \lambda_6 m2 + \lambda_{10} m_2 = 0, & \text{(A2a)} \\
(\sigma^1 \otimes \sigma^2): & \quad \lambda_3 m_2 + \lambda_5 m_1 - \lambda_9 m_6 + \lambda_{10} m_5 = 0, & \text{(A2b)} \\
(\sigma^2 \otimes \sigma^3): & \quad \lambda_2 m_4 + \lambda_3 m_6 + \lambda_7 m_3 - \lambda_9 m_2 = 0, & \text{(A2c)} \\
(\sigma^0 \otimes \sigma^2): & \quad \lambda_3 m_1 + \lambda_4 m_4 + \lambda_5 m_2 + \lambda_8 m_3 = 0, & \text{(A2d)} \\
(\sigma^2 \otimes \sigma^0): & \quad \lambda_1 m_5 + \lambda_2 m_6 + \lambda_3 m_4 + \lambda_4 m_1 = 0, & \text{(A2e)} \\
(\sigma^3 \otimes \sigma^2): & \quad \lambda_3 m_3 + \lambda_7 m_2 + \lambda_8 m_1 - \lambda_6 m_5 = 0. & \text{(A2f)}
\end{aligned}$$

Here we choose $\lambda_1 = \lambda_4 = \lambda_5 = \lambda_6 = \lambda_7 = \lambda_9 = 0$ to make $\boldsymbol{\Lambda}$ diagonal and the set of equations (A2) becomes

$$\lambda_3 m_5 + \lambda_{10} m_2 = 0, \tag{A3a}$$

$$\lambda_3 m_2 + \lambda_{10} m_5 = 0, \tag{A3b}$$

$$\lambda_2 m_4 + \lambda_3 m_6 = 0, \tag{A3c}$$

$$\lambda_3 m_1 + \lambda_8 m_3 = 0, \tag{A3d}$$

$$\lambda_2 m_6 + \lambda_3 m_4 = 0, \tag{A3e}$$

$$\lambda_3 m_3 + \lambda_8 m_1 = 0. \tag{A3f}$$

The mass matrices now reduce to

$$\boldsymbol{\Lambda} = \lambda_2(\sigma^0 \otimes \sigma^3) + \lambda_3(\sigma^0 \otimes \sigma^0) + \lambda_8(\sigma^3 \otimes \sigma^0) + \lambda_{10}(\sigma^3 \otimes \sigma^3), \tag{A4a}$$

$$\mathbf{m} = m_1(\sigma^0 \otimes \sigma^2) + m_2(\sigma^1 \otimes \sigma^2) + m_3(\sigma^3 \otimes \sigma^2)$$
$$+ m_4(\sigma^2 \otimes \sigma^0) + m_5(\sigma^2 \otimes \sigma^1) + m_6(\sigma^2 \otimes \sigma^3). \tag{A4b}$$

Now the $\boldsymbol{\Lambda}$ is diagonal. We first notice the diagonal part of $\mathbf{m}^2$ are

$$\mathbf{m}^2_{11} = (m_1 + m_3)^2 + (m_2 + m_5)^2 + (m_4 + m_6)^2, \tag{A5a}$$

$$\mathbf{m}^2_{22} = (m_1 + m_3)^2 + (m_2 - m_5)^2 + (m_4 - m_6)^2, \tag{A5b}$$

$$\mathbf{m}^2_{33} = (m_1 - m_3)^2 + (m_2 - m_5)^2 + (m_4 + m_6)^2, \tag{A5c}$$

$$\mathbf{m}^2_{44} = (m_1 - m_3)^2 + (m_2 + m_5)^2 + (m_4 - m_6)^2. \tag{A5d}$$

The requirement of off-diagonal terms in (4.1b) to vanish gives us the constraints on the choice of $m$'s:

$$m_1 m_2 = 0, \tag{A6a}$$

$$m_1 m_4 = 0, \tag{A6b}$$

$$m_5 m_4 = 0, \tag{A6c}$$

$$m_5 m_3 = 0, \tag{A6d}$$

$$m_6 m_2 = 0, \tag{A6e}$$

$$m_6 m_3 = 0. \tag{A6f}$$

First we notice that we have to choose at least three of the $m_i$'s to be zero to fulfill (A6):

$$m_1 = m_5 = m_6 = 0, \text{ or} \tag{A7a}$$

$$m_2 = m_3 = m_4 = 0. \tag{A7b}$$

However, these two choices lead us to nowhere since all $\lambda$'s have to be zero in order to satisfy the condition (A3).

Next we consider the case when four of the $m_i$'s are equal to zero. To satisfy (A6), only nine combinations are allowed:

$$m_1 = m_2 = m_3 = m_4 = 0, \tag{A8a}$$

$$m_1 = m_2 = m_3 = m_5 = 0, \tag{A8b}$$

$$m_1 = m_2 = m_5 = m_6 = 0, \tag{A8c}$$

$$m_1 = m_3 = m_4 = m_6 = 0, \tag{A8d}$$

$$m_1 = m_3 = m_5 = m_6 = 0, \tag{A8e}$$

$$m_1 = m_4 = m_5 = m_6 = 0, \tag{A8f}$$

$$m_2 = m_3 = m_4 = m_5 = 0, \tag{A8g}$$

$$m_2 = m_3 = m_4 = m_6 = 0, \tag{A8h}$$

$$m_2 = m_4 = m_5 = m_6 = 0. \tag{A8i}$$

Now let us stop the choosing process here for a while and back to (A5). A lesson from the two-flavor case is that the mode functions are both non-normalizable if $\mathbf{m}_{ii}^2 \neq 0$. Hence finding a single normalizable zero mode relies on $\mathbf{m}_{ii}^2 = 0$ and the signs of the eigenvalues of $\mathbf{\Lambda}$. From (A5), we see that the mass parameters come in pairs: $(m_1, m_3)$, $(m_2, m_5)$ and $(m_4, m_6)$ and they contribute to diagonal terms as $(m_i + m_j)^2$ to two of them and $(m_i - m_j)^2$ to the other two. Therefore, we conclude that we can only have single normalizable zero mode when the $m$'s vanish in pairs (with proper choice of $\mathbf{\Lambda}$). This reduces the possible choices of $m$'s (A8) to three:

$$m_1 = m_2 = m_3 = m_5 = 0, \tag{A9a}$$

$$m_1 = m_3 = m_4 = m_6 = 0, \tag{A9b}$$

$$m_2 = m_4 = m_5 = m_6 = 0. \tag{A9c}$$

Combing (A3) with (A9), we find the possible parameter choices for having only a single normalizable zero mode as follows:

$$m_1 = m_2 = m_3 = m_5 = 0, \ \lambda_2 = \lambda_3 = 0, \tag{A10a}$$

$$m_1 = m_3 = m_4 = m_6 = 0, \ \lambda_3 = \lambda_{10} = 0, \tag{A10b}$$

$$m_2 = m_4 = m_5 = m_6 = 0, \ \lambda_3 = \lambda_8 = 0. \tag{A10c}$$

Besides these three choices, we only have zero or two normalizable zero mode solutions. The corresponding $\mathbf{\Lambda}$ and $\mathbf{m}$ matrices for (A9) are:

$$\mathbf{\Lambda} = \begin{pmatrix} \lambda_{10} + \lambda_8 & 0 & 0 & 0 \\ 0 & \lambda_8 - \lambda_{10} & 0 & 0 \\ 0 & 0 & -(\lambda_8 + \lambda_{10}) & 0 \\ 0 & 0 & 0 & -(\lambda_8 - \lambda_{10}) \end{pmatrix},$$

$$\mathbf{m} = i \begin{pmatrix} 0 & 0 & -(m_4 + m_6) & 0 \\ 0 & 0 & 0 & -(m_4 - m_6) \\ m_4 + m_6 & 0 & 0 & 0 \\ 0 & m_4 - m_6 & 0 & 0 \end{pmatrix}, \quad \text{(A11a)}$$

$$\mathbf{\Lambda} = \begin{pmatrix} \lambda_2 + \lambda_8 & 0 & 0 & 0 \\ 0 & -(\lambda_2 - \lambda_8) & 0 & 0 \\ 0 & 0 & \lambda_2 - \lambda_8 & 0 \\ 0 & 0 & 0 & -(\lambda_2 + \lambda_8) \end{pmatrix},$$

$$\mathbf{m} = i \begin{pmatrix} 0 & 0 & 0 & -(m_2 + m_5) \\ 0 & 0 & m_2 - m_5 & 0 \\ 0 & -(m_2 - m_5) & 0 & 0 \\ m_2 + m_5 & 0 & 0 & 0 \end{pmatrix}, \quad \text{(A11b)}$$

$$\mathbf{\Lambda} = \begin{pmatrix} \lambda_2 + \lambda_{10} & 0 & 0 & 0 \\ 0 & \lambda_2 + \lambda_{10} & 0 & 0 \\ 0 & 0 & \lambda_2 - \lambda_{10} & 0 \\ 0 & 0 & 0 & -(\lambda_2 - \lambda_{10}) \end{pmatrix},$$

$$\mathbf{m} = i \begin{pmatrix} 0 & -(m_1 + m_3) & 0 & 0 \\ m_1 + m_3 & 0 & 0 & 0 \\ 0 & 0 & 0 & -(m_1 - m_3) \\ 0 & 0 & m_1 - m_3 & 0 \end{pmatrix}. \quad \text{(A11c)}$$